# Hysteresis phenomena in electron tunnelling, induced by surface plasmons

## Norbert Kroó*[1], Sándor Varró and Péter Rácz

Wigner Research Centre for Physics of the Hungarian Academy of Sciences
SZFI 1525 Budapest, P.O. Box 49, Hungary

Phone: + 36 1 392 2222/1027, Fax: + 36 1 331 4379
E-mail: norbert.kroo@office.mta.hu

[1]    * Corresponding author, E-mail: kroo.norbert@office.mta.hu

# Abstract

A high spatial resolution surface plasmon near field scanning tunnelling microscope (STM) has been used to study the properties of localized surface plasmons (SPO) in so-called hot spots on a gold surface, where the local electromagnetic field is extremely high. A CW semiconductor laser and a femtosecond Ti: Sa laser were used to excite the plasmons and the SPO excited tunnel current was used as the detector. When scanning the STM from negative to positive bias and reversed, hysteresis in the tunnel signal was found ,excluding (or rather minimizing) the role of the presence of a potential Casimir-effect in the process. It was found, however, that a multiple image charge induced double well potential may explain our experimental findings. The stepwise behaviour of the area of the observed hysteresis loops is a new, additional indication of the non-classical properties of SPO-s

**Keywords: Hysteresis,** scanning tunnelling microscope, surface plasmons, rectification,

## 1. Introduction

In recent years the number of papers on the study of surface plasmon oscillations (SPO) has dramatically increased. This phenomenon is connected partly with the unique properties of these collective excitations of conduction electrons on the surface of some metals, coupled to an evanescent electromagnetic field above the surface. These properties open up a broad spectrum of applications in information and communication technologies, chemistry, biology, medicine, etc. [1, 2].

One of these unique properties is the breaking of the diffraction limit, allowing to work with SPO-s in the nanoscale domain. This is a significant boost for optical applications, since the diffraction limit is one of the limiting factors in the use of "classical" optical methods in the nanoworld. An other feature of the near field of SPO-s is the concentration of the exciting electromagnetic field to a narrow sheet on the surface of the metal, resulting in field strengths a few ten times higher than that of the exciting laser field. Surface irregularities or metallic tips near the surface may localize the SPO-s, resulting in further field enhancement. In these high fields strong nonlinearities may occur and therefore SPO mediated nonlinear optical phenomena are observed at relatively low SPO exciting laser intensities [3]. There is an increasing number of applications which make use of these high fields

There is an other phenomenon which might be connected to SPO-s, namely the so-called Casimir effect [4, 5]. If two metallic plates are near to each other, the presence of quantum mechanical zero point fluctuations results in an attractive (or at larger distance repulsive) force. This is true for a surface-tip combination too. At small, nanometer range distances the force is attractive and is overwhelmingly an SPO mediated effect.

It has been found recently in several experiments that SPO-s have nonclassical properties. Among others, double slit interference or squeezing have been found [6]. Due to these exciting new developments a newly emerging field, namely quantum plasmonics is gaining in significance, with exciting new phenomena and potential applications.

In the present paper some of our new experimental findings are described in which the high spatial resolution of our SPO near field STM is explored. We concentrated to sites, where the plasmon field is extremely high and where the potential role of the Casimir-effect could be demonstrated. This idea has been raised, since at small distances between two metal plates, or surface-tip structures the role of SPO-s is significant. An attempt to give some qualitative explanations of our findings is also given. In Subsection 2.1 we describe the experimental setup, and present typical results on the hysteresis loops, found in the SPO-assisted voltage-current tunnelling signal. In Subsection 2.2 we are giving some semi-quantitative, or rather, qualitative interpretation of the measurements. In Section 3 a brief summary close our paper.

## 2. 1 Experimental apparatus and measurement results

Scanning tunnelling microscopes (STM) have been widely used to study the near field of surface plasmon oscillations (SPO) on metal surfaces with nanometer scale resolution [7]. In most cases SPO-s were resonantly excited in a thin gold film with lasers via a glass prism in the Kretschmann geometry, avoiding the direct heating of the STM tip by the laser beam [8]. When the SPO-s decay, they heat the metal film and the thermal dilatation is detected by the STM (thermal signal). If the material of the tip is different from that of the metal film an additional contact potential component is also added to the signal. It has however also been observed [9] that the EM field of SPO-s may be rectified, resulting in a signal which is fast compared to the thermal one (slow signal). In Figure 1. a typical, 30µs long exciting laser signal, a thermal and an SPO signal are shown. The STM can therefore be used to register simultaneously 3 images of the same surface, namely a topological, a thermal (slow) and an SPO near field (fast) image with high spatial resolution [10].

SPACE FOR FIGURE 1.

It was found, that the thermal signal image resembles to a certain extent the topological one, the SPO signal has however, a completely different distribution, localized mainly around grain boundaries of the surface. Figure 2 shows typical topographic and SPO images of a 100x100 nm gold surface. Here the gold film, evaporated onto the glass prism has been 45nm thick.

SPACE FOR FIGURE 2.

In some regions, between two grains on the surface, the so-called hot spots [11] the SPO signal has negative polarity (the gold surface has been grounded). In these hot surface regions the EM field of the localized SPO-s is enormous, in some cases around 1000 times stronger than that of the exciting laser field [10] and if the bias of the STM is scanned from negative to positive values, the SPO signal remains negative even at positive biases, up to a certain value (typically 40-100mV) where the field of the STM bias is equal to that of the rectified SPO field, and compensates it

The SPO-s were excited partly by a CW semiconductor laser (λ~670nm) which was gated to give 30µs pulses ( in Figure1. this pulse and the slow and fast response are shown) and by a Ti: Sa laser, giving ~ 100 fs, ~0.1μJ pulses, with 1.2 kHz frequency. The CW laser excitation resulted in both fast and slow signals while the fs ones gave only the fast ones.

It was found by us, that in cases, when the SPO signal was negative, even when the bias on the STM was zero, the negative SPO signal was still present [9]. A typical result of the STM bias scan from negative to positive values in shown in Fig.3 for both the fast and slow signals. During the scans the tip of the microscope was always kept at constant distance from the surface (equivalent to the distance of +10mV bias and 9pA tunnel current without SPO signals). The fast signal is negative, while the slow one is zero at zero bias. What we furthermore found, is that when decreasing the exciting laser intensity, even at low values, the negative SPO signal is there and if extrapolated to zero bias value it is still different from zero (inset of Fig 3).

SPACE FOR FIGURE 3

In order to find out the physical background of this observation two potential explanations may be considered. The first one is the  so-called Casimir effect, resulting in a force between the tip and the gold surface due to the wide spectrum of quantum mechanical zero point fluctuations [4 ]. The second one is the decrease of the height of the potential barrier between tip and surface due to the formation of a multiple image charge induced double well potential, leading to tunnelling at lower biases.

In the first case the effect should be independent from the polarity of the tip, while in the second one it should not. Therefore we performed negative – positive – negative scans of our STM. The basic (no SPO signal) and the thermal signals showed no hysteresis during these up-and-down scans, while the SPO signal did. A typical hysteresis curve is shown in Fig 4.

The conclusion is, that the Casimir effect can be excluded (or at least that it's contribution is small) and the potential explanation is the multiple mirroring effect, namely the decrease of the potential barrier between tip and surface due to this phenomenon. See a more detailed explanation in Subsection 2.2.

SPACE FOR FIGURE 4

The measurements were carried out also with the same microscope and gold film, but with femtosecond (100-120fs) pulsed laser excitation. The same hysteresis has here also ben found.A typical hysteresis curve is shown in Fig.5, while the inset of this figure shows the response SPO signal of the STM to a single laser pulse. Therefore the pulsed laser experiments confirm our conclusions, drawn from our CW laser observations.

When measuring numerous hysteresis loops at different hot spots on the gold surface, in both sets of experiments, a somewhat surprising result has been found. And the same observation is true for measurements at different laser intensities. Namely, when the area of measured loops were plotted in increasing hysteresis loop area order, a stepwise plot was found The results of the pulsed laser excitation case are shown in Fig 5., but similar result was found in the CW laser excitation case too. The single steps are clearly seen and plotted in the inset of this figure. This observation is considered as an additional to [6,7 ] proof of the non-classical behaviour of SPO-s. It was also found, that if the hysteresis loops were cut at a fixed STM bias, the difference between the up and down scan signals show also this step-like behaviour.

SPACE FOR FIGURE 5

## 2. 2 Possible qualitative and quantitative interpretations of the experimental results

In principle, due to the field-enhancement in the SPOs [12] one would also expect a direct photon absorption mechanism taking place during the laser pulses, such that the detection of a tunnel current component depends nonlinearly on the laser intensity [13], we do not think that this would be a sizable effect under our experimental conditions. It is also unlikely that the tunnelling signals are coupled to some discrete phonon or vibrational excitations [14] due to a transient heating, which may be responsible for the manifestly discrete structure of the loop area shown in Fig.6. The hysteresis found by us, and illustrated in Figs. 4-5, suggest the assumption of some nonlinear mechanism resulting in a bistable dynamics of the tunnel electrons. We note at this point that quite recently optical bistability has been found in the optical transmission through nanostructures, and this effect has been explained by "the strong sensitivity of the surface-plasmon mode resonances at the signal wavelength to the surrounding dielectric environment and the electromagnetic field enhancement due to plasmonic excitations at the controlled light wavelengths" [15]. A more closely related subject is the study of the switching characteristics of nanoscale junctions created between a tungsten tip and a silver film covered by a thin ionic conductor layer [16]. In this quoted study it has been shown that the atomic-sized junctions the switching vary randomly for different junctions [16]. Accordingly, in our experiments it is natural to expect an inherent randomness of the tunnel current, because of the very small (~ atomic) sizes of the closest-approach areas of both the tip and of the localized SPOs generated by the laser pulses. However, to our knowledge, no such kind of quantitative investigations has been carried out so far. It has recently been shown that, in the presence of an external magnetic field, genuine single-particle quantum interference phenomena may cause an asymmetric change in the density of states [17]. In principle, an 'anholonomy' in the density od states may also be responsible for a hysteresis phenomenon in the STM current, because this depends on the density of states, too. By now, we have not been able to give a complete interpretation of the measured phenomena, except for the linear intensity dependence of the current.

The following analysis will rely on a rather simple description, which is based on the elementary theory of tunnelling [18] and on the image charge picture and formulae of classical electrodynamics. In our view, the hysteresis found in our experiments may perhaps be related to the bistable dynamics of the electron passage between the tip and the sample, and to the inherent randomness of the current. Our starting point is to take into account that, down to the spacing of order 3 Angströms, the (multiple) image potential picture of classical electrodynamics is still applicable. If the tip and the sample are such close to each other, as in the usual STM experiments, then even at zero bias, the multiple image charge may result in a considerable reduction of the barrier between the anode and cathode [18]. It seems to us that the possible role of this multiple image has not received that attention as it deserves. This potential is described by the digamma function $\psi(x)$, which is often called simply 'psi–function' [19].

$$V_{im}(z) = -\frac{e^2}{2d}\left[\psi(1) - \frac{1}{2}\psi(z/d) - \frac{1}{2}\psi(1 - z/d)\right], \quad \psi(x) = \frac{1}{\Gamma(x)}\frac{d\Gamma(x)}{dx}, \qquad (1)$$

where $e$ is the electron's charge and $d$ is the distance between the tip and the sample surface. Strictly speaking, the above formula is valid only for a test charge between two ideal metal plane surfaces (see Fig. 7), however, its transverse variation can also be taken into account by an approximate use of parabolic coordinates

$$V_{im}(z,r) \approx -\frac{e^2}{2d}\left[\psi(1) - \frac{1}{2}\psi(\xi/d) - \frac{1}{2}\psi(1 - \xi/d)\right], \quad \xi \equiv [z + (z^2 + r^2)^{1/2}]/2, \qquad (2)$$

where $z$ denotes the longitudinal coordinate along the tip axis, and $r$ measures the offset perpendicular to the it. The distortion of the potential due to the curvature at the tip end is illustrated in Fig. 8. It is clear that in the presence of an applied bias, the above potential becomes assymmetric longitudinally. The maximum value of this image potential energy of a test charge between the two metal surfaces, is the universal function of the tip-sample distance;

$$-V_{im}^{max} = -V_{im}(d/2) = \frac{e^2}{2d}ln2 = \frac{0.9427}{[d/1nm]}eV, \qquad (3)$$

and the vacuum level of the original potential may be considerably broken down if the tip-sample distance is small (as is illustrated in Fig. 9).

SPACE FOR FIGURE 7

SPACE FOR FIGURE 8

SPACE FOR FIGURE 9

In fact, the presence of the multiple image charge results in the formation of the double-well potential between the tip and the sample (see Eq. (1) and Fig. 7), and the enhanced field of the surface plasmons induce the hopping between the two minima. Since the current is inherently random at such small distances, the jumps in the opposite directions are necessarily not symmetric, as one varies the bias voltage. The transmission coefficient $D(v_z)$, as a function of the electron's longitudinal velocity $v_z$, behaves like [18]

$$D(v_z) = 1 \text{ for } v_z > \sqrt{\frac{2}{m}\left(V_0 - \frac{e^2}{d}ln2\right)} \equiv v_0, \text{ and } D(v_z) \text{ decays exponentially for } v_z < v_0, \quad (4)$$

where $V_0$ is the asyptotic potential difference between vacuum and the inner-metal potential in the conduction band. By taking Eq. (4) into account [18], the relevant part of the current density can be expressed as

$$J = 4\pi\frac{m^2}{h^3}\Delta\int_{\sqrt{2A/m}}^{\sqrt{2V_0/m}}exp[-2mud/\hbar]udu = 4\pi\frac{m^2}{h^3}\left((e^2/d)ln2 - A\right)\Delta, \qquad (5)$$

where $A$ and $\Delta$ are the effective work function and the total bias, respectively, and $m$ is the electron's mass, $h$ denotes Planck's constant. The time-average action of the localized SPOs can be represented by the effective bias stemming from the gradient of its ponderomotive potential energy, $U_p(\boldsymbol{r}) = e^2E^2(\boldsymbol{r})/4m\omega^2$, where $E(\boldsymbol{r})$ is the amplitude distribution of the electric field strength of the SPOs. Since this (enhanced) field is simply proportional with the incoming laser intensity, we conclude that $\Delta \sim |gradU_p(\boldsymbol{r})| \sim I$, i.e. the effective bias is proportional with the incoming laser intensity. Accordingly, by taking Eq. (5) into account, we have $J \sim I$, in agreement with our experimental results. Thus, we have given a possible interpretation of the measured essentially linear intensity-dependence of the SPO induced tunnelling on the incoming laser intensity. This is a seemingly paradoxical result, because usually some nonlinear mechanisms are supposed to lead to the rectification phenomena. Of course, the detailed dependence is more complex, since the local distance $d$ also depends on the intensity due the local warming up of the components.

## 3. Summary

Our STM was scanned by changing the bias through negative and positive values. Some of the typical experimental results are presented in Subsection 2.1.The fast (SPO) signals of the STM (detected in hot spots) were always found to be negative even at zero bias, in contrast to the slow (thermal) signals, being always zero at zero bias. These negative signals showed hysteresis in negative – positive – negative bias scans, while the slow signals did not. This feature was found in both cases, when the SPO-s were excited by a CW and by a femtosecond pulsed laser. On the basis of the experimental results the Casimir-effect as the potential explanation of the nonzero SPO signal at zero bias of the STM could be excluded. When plotting the area of the numerous hysteresis loops (measured at different hot spots) in increasing area sequence, a stepwise function was found. The stepwise function has been found not only for loop areas, measured at different hot spots, but similarly at fixed bias values, plotting the difference between up and down scan signals

In Subsection 2.2 we have given a qualitative explanation of the hysteresis phenomena on the basis of the double-well potential, induced by the multiple image charges when the tip-sample distance is very small. Besides, we have also given a possible interpretation of the measured, essentially linear intensity-dependence of the SPO induced tunnelling current as the function of the incoming laser intensity.

The stepped function of Fig.5 is a further indication to the non-classical properties of SPO-s, but the detailed experimental and theoretical explanations need further studies.

**Acknowledgements:** This work has been supported by grants of the Hungarian Academy of Sciences and the Hungarian National Scientific Research Foundation OTKA, (Grant No. K73728). The authors thank P.Dombi for the use of the fs laser.

## References

[1] A. V. Zayats, I. I. Somolyaninov, A. A. Maradudin, *Physics Reports* **408** (2005) 131-314

[2] N. Halas, S. Lal, W. S. Chang, S. Link, P. Nordlander, *Chem. Reviews* **111** (2011) 3913

[3] N. Kroo, S. Varro, G. Farkas, P. Dombi, D. Oszetzky, A. Nagy, A. Czitrovszky,
*Journal of Modern Optics* **55** (2008) 3203-3210

[4] H. B. G. Casimir, *Proc. Kon. Ned. Acad. Watenschap* **51**, 793 (1935)

[5] F. Intravaley, A. Lambrecht, *Phys. Rev. Lett.* **94** (2005) 110404

[6] N. Kroo, S. Varro, G. Farkas, D. Oszetzky, A. Nagy, A. Czitrovszky, *Journal of Modern Optics* **54** (2007) 2679-2688

[7] N. Kroo, Z. Szentirmay, H. Walther, *Surf. Sci.* **582** (2005) 110

[8] E. Kretschmann, H. Raether, *Z. Naturforsch.* A**23** (1968) 2135

[9] N. Kroo, J. P. Thost, M. Völcker, W. Krieger, H. Walther, *Europhys. Lett*. **15** (1991) 289

[10] M. Lenner, P. Rácz, P. Dombi, G. Farkas, N. Kroo, *Phys. Rev. B* **83** (2011) 205428

[11] K. Li, M. J. Stockman, D. J. Bergman, *Phys. Rev. Lett.* **91** (2003) 227402

[12] S. Varró, N. Kroó, Gy. Farkas and P. Dombi, *J. Mod. Opt*., **57**, (2010) 80-90

[13] M. Merschdorf, W. Pfeiffer, A. Thon and G. Gerber, *Appl. Phys. Lett*. **81**, (2002) 286-288

[14] A. N. Pasupathy, J. Park, C. Chang et al., *Nano Letters* **5**, (2005) 203-207

[15] G. A. Wurtz, R. Pollard, and A.V. Zayats, *Phys. Rev. Lett*. **97,** (2006) 057402

[16] A. Geresdi, A. Halbritter, A. Gyenis, P. Makk, and G. Mihály, *arXiv:* 1006.5402v3 [cond-mat.mes-hall] (22 Apr 2011).

[17] A. Cano and I. Paul, *Phys. Rev. B* **80**, (2009) 153401

[18] A. Sommerfeld u. H. Bethe, *Elektronentheorie der Metalle* (Springer, Berlin, 1967)

[19] I. S. Gradshteyn and I. M. Ryzhik, Table of Integrals, Series, and Products. (Academic Press, New York, 1980)

**Figures and captions to figures.**

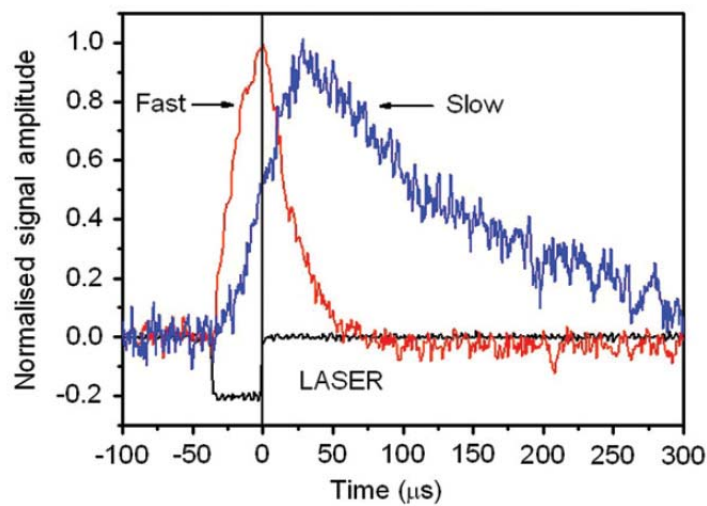

**Fig. 1** Time dependence of the chopped laser, exciting the SPO-s in the Kretshmann geometry and typical slow (thermal) and fast (SPO) response signals of the scanning tunnelling microscope (STM).

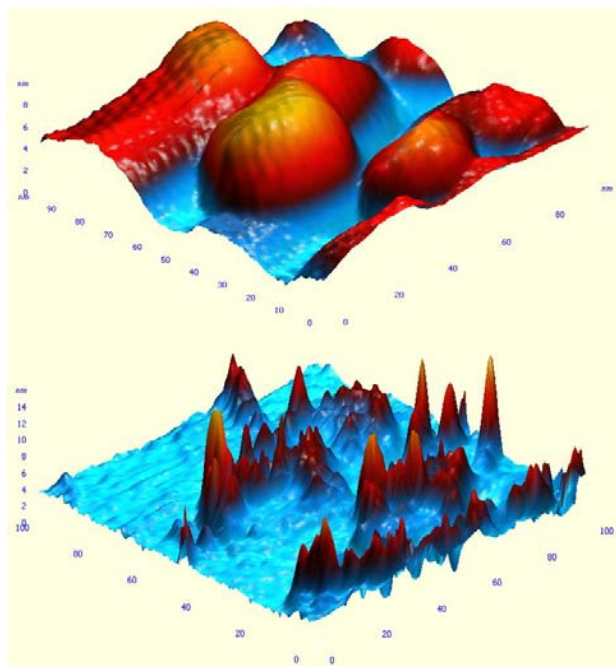

**Fig. 2** (a) upper figure: Topography image of a 100x100nm gold surface. (b) lower figure: SPO image of the same surface. The SPO-s are localized around grain boundaries where the EM field is extremely high. In some points (hot spots) the EM field of SPO-s is not only rectified, but negative, independent of the polarity of the bias on the STM, up to a certain positive value.

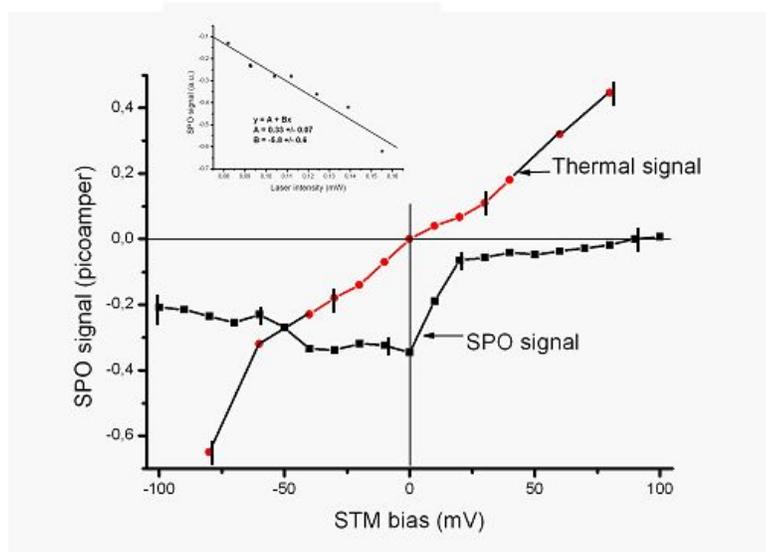

**Fig. 3** The STM bias is scanned from negative to positive values. The thermal signal changes sign together with the bias and is always zero at zero STM bias. The SPO signal at "negative" hot spots  remains negative at zero STM bias and even above. Inset: The laser intensity dependence of the SPO signal at zero STM bias. When extrapolated to zero laser intensity the signal is still different from zero.

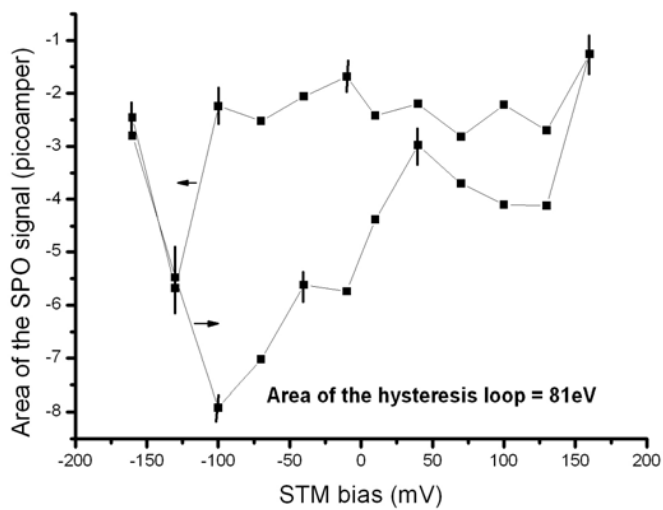

**Fig. 4** Hysteresis loop of the SPO signal. The scan started and finished at negative bias values. Each point in the figure represents the area of the total SPO signals in electronvolts ([signal(amper) x pulse duration(sec) x bias of the STM(volt)] = Vas = Wattsec → eV).

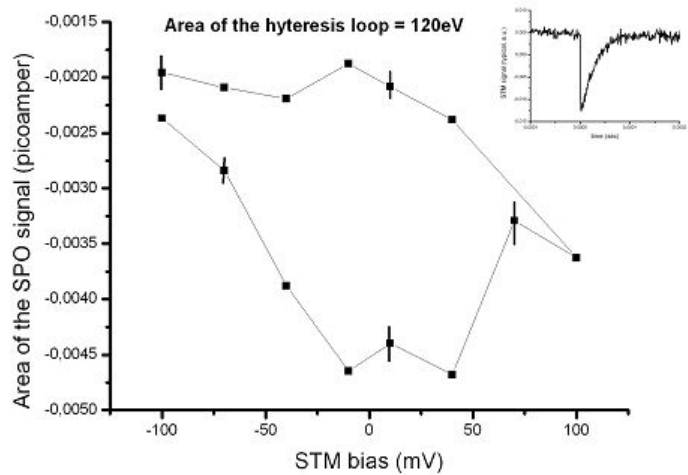

**Fig. 5** The SPO near field is excited in the microscope by a 100-120 fs long, 1.2 kHz frequency Ti: Sa Laser (λ=800nm). Similar hysteresis loops have been found as in the CW laser case (as in Fig.4). Inset: A typical negative signal (Δt=28μs) is shown. It is the result of a single fs laser pulse excitation (laser intensity: ~300mW).

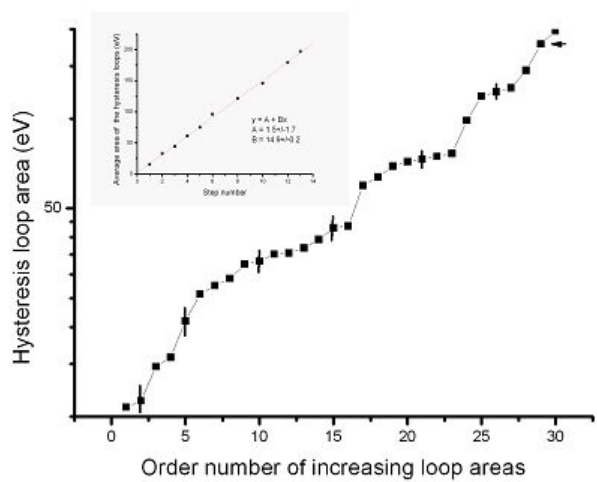

**Fig. 6** A large number of loops were measured with both types of laser excitations. Here the fs laser results are shown. The hysteresis loop areas are plotted in increasing area order. Discrete steps are found, indicating non-classical properties of the SPO-s. Inset: Plot of the average areas of consecutive, equidistant steps.

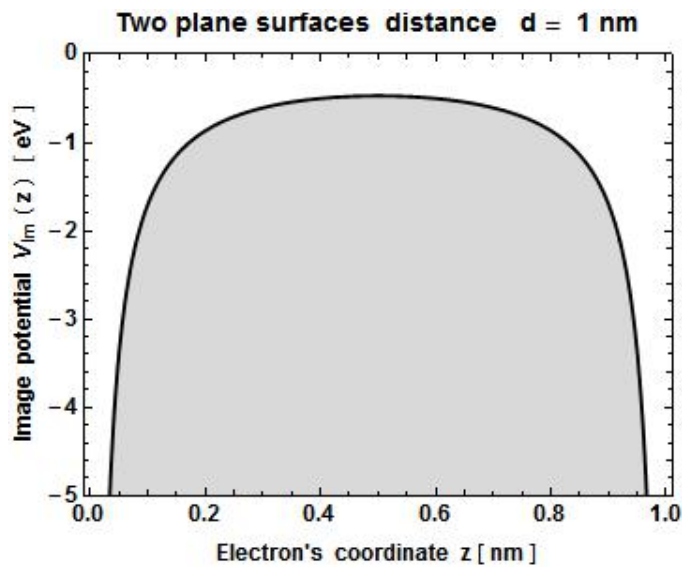

**Fig. 7** Shows, the image potential between the two metal surfaces which stems from the multiple reflection of the test charge at zero bias. This universal curve has been drawn on the basis of Eq. (1).

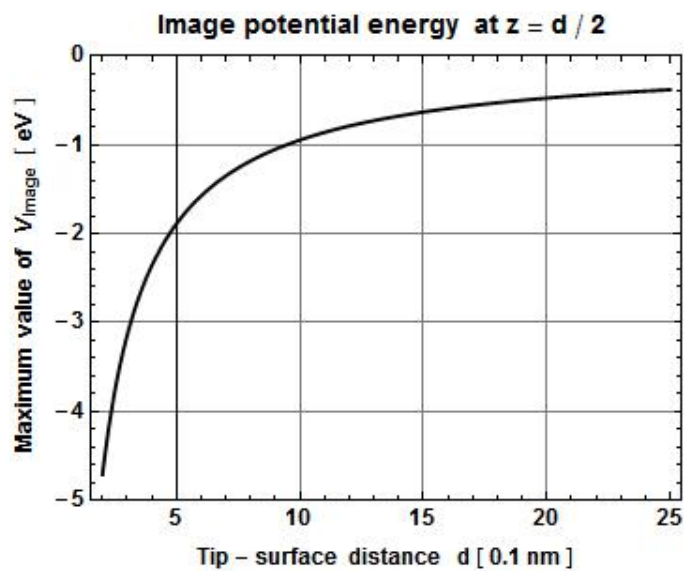

**Fig. 8** Shows the maximum value of the image potential energy of a test charge between the two metal surfaces, according to the formula given by Eq. (2). This universal functional dependence clearly shows that the breakdown of the vacuum level of the potential can be very large if the tip-sample distance is small.

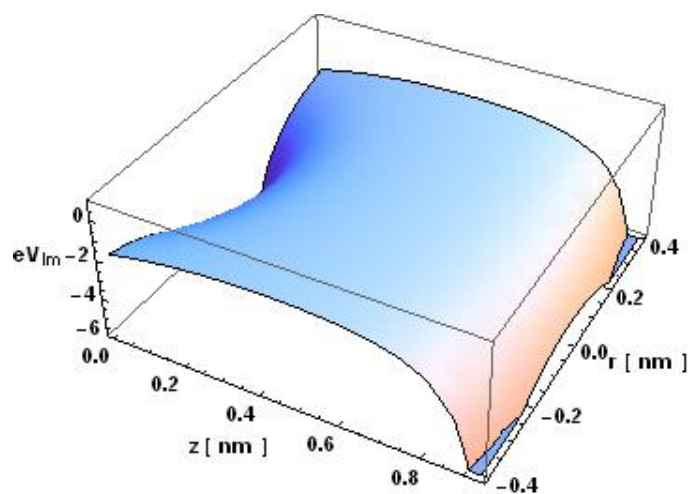

**Fig. 9** Illustrates the approximate 3-dimensional image potential energy of a test charge between the curved tip surface and an ideal plane sample metal surfaces. The approximation formula given by Eq. (3), has been used, where the radial coordinate $r$ is measured from the centre of the tip end. The tip's shape has been assumed to be a paraboloid of revolution. We emphasize that here we did not use the exact solution of the electrostatic problem, but, rather we simply have 'put by hand' the parabolic coordinate $\xi$ to the original planar formula of Eq. (1), just for illustration purposes.